\documentclass[runningheads]{llncs}
\let\llncssubparagraph\subparagraph
\let\subparagraph\paragraph
\usepackage[compact]{titlesec}
\let\subparagraph\llncssubparagraph

\usepackage{graphicx}
\usepackage{color}
\usepackage{multirow}
\usepackage{booktabs}
\usepackage{caption}
\usepackage{subcaption}
\usepackage{titlesec}

\usepackage[belowskip=2pt,aboveskip=2pt]{caption}

\titlespacing{\section}{0pt}{2pt}{5pt}
\titlespacing{\subsection}{0pt}{2pt}{5pt}

\begin{document}
\title{On Analyzing Antisocial Behaviors Amid COVID-19 Pandemic}
%
%
\author{Md Rabiul Awal,\textsuperscript{1} Rui Cao,\textsuperscript{2} Sandra Mitrovi\'c,\textsuperscript{3} Roy Ka-Wei Lee\textsuperscript{1}}


\institute{University of Saskatchewan, Saskatoon, SK, Canada \and
University of Electronic Science and Technology of China, Chengdu, Sichuan, China \and
Istituto Dalle Molle di Studi sull'Intelligenza Artificiale, Switzerland\\
\email{\{mda219,roylee\}@cs.usask.ca, caorui0503@gmail.com, sandra.mitrovic@idsia.ch}}

\maketitle              
\begin{abstract}
The COVID-19 pandemic has developed to be more than a bio-crisis as global news has reported a sharp rise in xenophobia and discrimination in both online and offline communities. Such toxic behaviors take a heavy toll on society, especially during these daunting times. Despite the gravity of the issue, very few studies have studied online antisocial behaviors amid the COVID-19 pandemic. In this paper, we fill the research gap by collecting and annotating a large dataset of over 40 million COVID-19 related tweets. Specially, we propose an annotation framework to annotate the antisocial behavior tweets automatically. We also conduct an empirical analysis of our annotated dataset and found that new abusive lexicons are introduced amid the COVID-19 pandemic. Our study also identified the vulnerable targets of antisocial behaviors and the factors that influence the spreading of online antisocial content.

\keywords{Antisocial Behavior  \and Pandemic \and Social Media.}
\end{abstract}
\section{Introduction}
\label{sec:introduction}
The COVID-19 pandemic, which is the ongoing pandemic of coronavirus disease 2019 (COVID-19) caused by the severe acute respiratory syndrome coronavirus 2 (SARS-CoV-2) \cite{mehta2020covid}, has seen more than 5 million reported diagnosed cases and more than 300,000 deaths globally as of May 2020. Globally, countries have taken unprecedented measures such as social distancing, contact tracing, border closures, etc., to curb the spread of the coronavirus. However, the COVID-19 pandemic has developed to be more than a bio-crisis as global news has reported a sharp rise in xenophobia and discrimination in both online and offline communities \cite{devakumar2020racism,unnews2020}. The hostility and discrimination towards Chinese communities in online social media have particularly increased as China was presumed to be the ground zero of the coronavirus \cite{schild2020go}. 

The spread of online antisocial behaviors is not unique to the COVID-19 pandemic. The analysis and detection of online antisocial behaviors such as hate speeches \cite{schmidt2017survey,fortuna2018survey,waseem2016you,waseem-hovy:2016:N16-2,davidson2017automated,founta2018large,deGibert2018,DBLP:conf/icwsm/ElSheriefKNWB18,arango2019hate,FountaCKBVL19}, cyberbullying \cite{chatzakou2017mean,agrawal2018deep}, harassment \cite{golbeck2017large}, and usage of abusive and offensive languages \cite{davidson2017automated,waseem-hovy:2016:N16-2,FountaCKBVL19,zampieri2019predicting,zampieri2019semeval}, etc., have been widely studied by the data mining research community. The spread of antisocial behaviors in social media has not only sowed discord among individuals or communities online but also resulted in violent hate crimes \cite{Williams19,relia2019race,mathew2019spread}. Therefore, it is a pressing issue to detect and curb online antisocial behaviors, particularly during a pandemic, as preventing such undesirable online behaviors can avoid adding problems to the already difficult crisis.

Despite the gravity of the issue, the literature addressing online antisocial behaviors amid the COVID-19 pandemic is scarce. Schild et al. \cite{schild2020go} performed an empirical analysis of the emergence of Sinophobic behavior on the Web during the outbreak of the COVID-19 pandemic. The researchers focused their analysis on data collected from Twitter and 4Chan and found that there was a rise of Sinophobic behaviors amid the COVID-19 period. While the study did provide some preliminary understanding of the spread of Sinophobic behaviors during the COVID-19 pandemic, it was limited to only analyzing the discrimination towards the Chinese community. There could be online antisocial behaviors targeting other vulnerable individuals and groups amid the pandemic that are neglected in the existing study, and our study aims to fill this research gap. 

A key challenge to the study on online antisocial behaviors amid the COVID-19 pandemic is the lack of annotated datasets. To our best knowledge, there is no publicly available annotated online antisocial behavior dataset in the context of COVID-19. In this paper, we bridge the gap by first collecting a large dataset containing more than 40 million social media posts related to the COVID-19 pandemic. Ideally, we would manually examine and annotate the antisocial behaviors in the collected dataset. However, such an approach is too laborious and time-consuming as we have collected a large dataset. Therefore, we propose a novel framework to annotate the antisocial behaviors in the dataset automatically. We adopted two main methods to annotate the dataset: (1) a lexicon-based approach, which performs keywords or key-phrases matching to automatically detect social media posts with antisocial slurs, and (2) the Google and Jigsaw's \textit{Perspective} API\footnote{https://www.perspectiveapi.com/}, which is an open-source program that analyzes the toxicity in a given text. Finally, the annotations from the two methods are combined to generate the final annotated antisocial behavior dataset. We will conduct a preliminary empirical analysis on the annotated dataset to provide some insights on the online antisocial behaviors amid the COVID-19 pandemic. The large annotated dataset will also be released in the hope that it will foster and encourage more research in this important area. 

We summarize our contributions in this paper as follows:

\begin{enumerate}
    \item We propose an automatic online antisocial behavior annotation framework to annotate one of the largest antisocial behavior datasets collected amid a pandemic event. 
    \item We collect and release a large annotated online antisocial behavior dataset to enable and encourage more research in online antisocial behaviors. 
    \item We provide preliminary empirical analysis on the online antisocial behaviors amid the COVID-19 pandemic.
\end{enumerate}

The rest of the paper is organized as follows: We first review the related literature in Section \ref{sec:related}. Section \ref{sec:annotation} present the data collection process and our proposed automatic online antisocial behavior annotation framework. We discuss our preliminary empirical analysis on the online antisocial behaviors amid the COVID-19 pandemic in Section \ref{sec:empirical}. Finally, we conclude the paper in Section \ref{sec:conclusion}.


\section{Related Work}
\label{sec:related}


Analyzing and detecting online antisocial behavior is a widely studied research area. In this section, we survey two groups of literature relevant to our paper: (1) data collection and annotation of online antisocial behaviors, and (2) other social media datasets collected amid the COVID-19 pandemic.


Data collection and annotation of online antisocial behaviors is an essential process in all online antisocial behaviors studies. Existing studies have collected social media datasets to analyze online antisocial behaviors such as hate speech \cite{waseem2016you,waseem-hovy:2016:N16-2,davidson2017automated,founta2018large,deGibert2018,DBLP:conf/icwsm/ElSheriefKNWB18,arango2019hate,FountaCKBVL19}, harassment \cite{golbeck2017large}, cyberbullying \cite{chatzakou2017mean,agrawal2018deep}, and usage of offensive and abusive languages \cite{davidson2017automated,waseem-hovy:2016:N16-2,FountaCKBVL19,zampieri2019predicting,zampieri2019semeval,waseem-hovy:2016:N16-2,founta2018large}. The annotation of antisocial content in social media is a difficult task. A common approach is to recruit manual annotators \cite{waseem-hovy:2016:N16-2,golbeck2017large,deGibert2018} independently or via crowdsourcing platforms such as Figure Eight \cite{zampieri2019predicting} and CrowdFlower \cite{chatzakou2017mean,founta2018large,davidson2017automated,DBLP:conf/icwsm/ElSheriefKNWB18}. However, manual annotation is expensive, labour-intensive, and time-consuming. Particularly for large datasets, it is impractical to annotate all the social media posts. Furthermore, manual annotation is susceptible to the annotators' bias \cite{waseem2016you}.

To overcome the limitations of manual annotations, studies have explored lexicon-based methods to complement the manual annotation process or annotate the antisocial behaviors fully automatically \cite{waseem-hovy:2016:N16-2,golbeck2017large,davidson2017automated,DBLP:conf/icwsm/ElSheriefKNWB18,qian2019benchmark}. Commonly, a lexicon containing antisocial keywords are first identified. Subsequently, keywords matching will be performed to annotate social media posts by checking if the textual content contains antisocial keywords. However, the lexicon-based methods also have limitations. Due to the informal nature of social media, the textual content is often short and contains grammatical errors, making it difficult for lexicon-based methods to perform keywords matching. Furthermore, many keywords can be used in both appropriate and antisocial contexts, and new antisocial lexicons may be developed over time \cite{davidson2017automated}. In this paper, we proposed an annotation framework that applies an open-source content toxicity analysis API to complement the lexicon-based approach.

The outbreak of the COVID-19 pandemic has motivated the spike increase in medical research and social media studies relevant to the pandemic. Gao et al. \cite{gao2020naist} collected COVID-19 related posts from Twitter and Weibo and performed preliminary keyword analysis on the two social media platforms. Other studies have also similarly collected large-scale COVID-19 multilingual Twitter datasets \cite{banda2020large,chen2020covid,lopez2020understanding,kabir2020coronavis}. However, most of these datasets are for open scientific research, and they are not annotated for antisocial behavior research. The closest related work is the study proposed by Schild et al. \cite{schild2020go}. The researchers performed an empirical analysis of the emergence of Sinophobic behavior on Twitter and 4Chan during the outbreak of the COVID-19 pandemic. While the study did provide some preliminary understanding of the spread of Sinophobic behaviors during the COVID-19 pandemic, it neglects other online antisocial behaviors targeting other vulnerable individuals and groups amid the pandemic. Furthermore, the study adopted an empirical analysis approach and did not perform any annotations on the dataset. Our study aims to fill this research gap by annotating the online antisocial behaviors in a COVID-19 related dataset and perform a holistic analysis of the antisocial behaviors amid the pandemic.

\section{Antisocial Behavior Annotation Framework}
\label{sec:annotation}

In this section, we first describe our data collection process, where we collected a large Twitter dataset comprising of tweets related to the COVID-19 pandemic. Next, we present our proposed antisocial behavior annotation framework. Specifically, we discuss the two main methods included in our framework: the lexicon-based method and the \textit{Perspective} API.   

\subsection{Data Collection}
We focus our data collection efforts on the Twitter platform. To retrieve a set of COVID-19 related tweets, we first define a set of case-sensitive keywords that are frequently used in the COVID-19 news and discussions. These keywords include: ``covid-19'', ``COVID-19'', ``COVID'', ``Coronavirus'', ``coronavirus'', ``CoronaVirus'', and ``corona''. Next, we utilize Twitter's Streaming API to retrieve a set of COVID-19 related tweets using the earlier defined keywords as queries. The comments on these retrieved tweets are also collected. For this study, tweets from 17 March 2020 to 28 April 2020 are collected. We further filter and remove the non-English tweets in our collected dataset, resulting in a total of 40,385,257 tweets retrieved. We term final collected dataset as the \textit{COVID-19 dataset}.


\subsection{Antisocial Behavior Annotation Framework}
The COVID-19 dataset is considerably large compared with existing publicly available antisocial online behavior datasets~\cite{waseem-hovy:2016:N16-2,DBLP:conf/acl-alw/ParkF17,DBLP:conf/icwsm/DavidsonWMW17,DBLP:conf/icwsm/FountaDCLBSVSK18}. Therefore, it is impractical to annotate the COVID-19 dataset manually. We propose an annotation framework to annotate the antisocial behavior in the COVID-19 dataset automatically. While we agree with existing studies that there could be many sub-categories of antisocial behaviors \cite{waseem-hovy:2016:N16-2,DBLP:conf/icwsm/DavidsonWMW17,DBLP:conf/icwsm/FountaDCLBSVSK18}, it is particularly challenging to annotate antisocial behaviors at fine-grain level automatically. Instead, we simplified the annotation process by annotating the tweets with binary labels: ``normal'' and ``antisocial''. Our proposed antisocial behavior annotation framework mainly comprises two annotation techniques: the lexicon-based method and the \textit{Perspective} API.

\textbf{Lexicon-Based Method.} We first compile a word corpus of antisocial keywords from various open-source antisocial behavior and online toxic content lexicons: HateBase\footnote{The largest structured hate speech repository, available at https://hatebase.org}, RSBD\footnote{http://www.rsdb.org/}, and Wikipedia\footnote{https://en.wikipedia.org/wiki/List\_of\_ethnic\_slurs}. Next, we manually remove ambiguous words such as ``pancake'', ``yellow
'', etc., as these words could also be used in normal conversational settings. As the word corpus may contain rare slurs that are obsoleted or no longer relevant, we further filter the word corpus by checking its relevance to social media. To perform this operation, we first construct a combined annotated antisocial behavior dataset by aggregating antisocial content from three publicly available datasets: WZ-LS~\cite{DBLP:conf/acl-alw/ParkF17}, DT~\cite{DBLP:conf/icwsm/DavidsonWMW17} and FOUNTA~\cite{DBLP:conf/icwsm/FountaDCLBSVSK18}. Subsequently, we perform frequency count for each keyword in the antisocial word corpus by computing the number of times it occurs in the combined annotated antisocial behavior dataset. Finally, infrequent antisocial keywords that occurred less than five times are removed from the antisocial word corpus. We term this final antisocial word corpus from open-source lexicons as the \textit{basic lexicon set}.

We also noted that there might be words that share similar antisocial semantic as the keywords in the \textit{basic lexicon set} but are not included in the lexicon itself. To this end, we construct an \textit{extended lexicon set}, which include keywords with similar antisocial semantic. We first train a Word2Vec model \cite{mikolov2013distributed} over the COVID-19 dataset. Next, we search keywords that share similar semantics with the keywords in the \textit{basic lexicon set}. More specifically, keywords that have more than 0.7 similarity scores with any keywords in the \textit{basic lexicon set} are included in the \textit{extended lexicon set}. 

Finally, we annotate the COVID-19 dataset by performing keyword matching against the \textit{basic lexicon set} and \textit{extended lexicon set}. Specifically, tweets that contain any keywords in the \textit{basic lexicon set} and \textit{extended lexicon set} are labeled as ``antisocial'', while the rest of the tweets are deemed as ``normal''. 


\textbf{\textit{Perspective} API\footnote{https://www.perspectiveapi.com/}} While the lexicon-based method is simple and would be able to identify a substantial amount of antisocial behaviors online, it still has limitations. For instance, there might be new antisocial keywords that are not included in the open-source antisocial word corpus. To address this limitation, we added another automatic annotation approach. The \textit{Perspective} API is an open-source program developed by Google's Counter Abuse Technology team and Jigsaw in order to improve online discussions. The API scores a given text based on several categories, such as toxicity, profanity, insult, etc. For each of these categories, Perspective API trains classifiers and outputs the probability score of a given text with respect to specific categories. Among the categories, the \textit{toxicity} score aligned most to our annotation goal. Therefore, we use the toxicity score in the \textit{Perspective} API to annotate the COVID-19 dataset. Specifically, we label a tweet as ``antisocial'' when it is given a $>0.5$ toxicity score by the \textit{Perspective} API, otherwise, the tweet will be labeled as ``normal''.


 

\begin{table}[t!]
  \caption{Result for the automatic annotation of the COVID-19 dataset.}
  \label{tab:anno-result}
  \centering
  \begin{tabular}{|c|c|c|c}
    \hline
    \textbf{Method} & \textbf{Antisocial tweets} & \textbf{Normal tweets} \\\hline\hline
    Lexicon-based & 1,169,755 & 39,215,502 \\\hline
    Perspective API & 2,383,316 & 38,001,941\\\hline
    Combined & 2,659,585 & 37,725,672\\\hline
\end{tabular}
\end{table}


As each annotation method has its strengths, we combine the lexicon-based method and Perspective API to annotate our COVID-19 dataset. We annotate a tweet as ``antisocial'' if it is a annotated as such by any of the two methods. Otherwise, the tweet will be labeled as ``Normal''. The subsequent empirical analysis will be based on our annotated COVID-19 dataset\footnote{Due to double blind policy, the link to the dataset will be released after paper acceptance}. Table \ref{tab:anno-result} shows our final annotation results, where about 7\% of the tweets are annotated to contain antisocial content.

\section{Empirical Analysis}
\label{sec:empirical}

In this section, we perform preliminary empirical analysis on the annotated antisocial behavior COVID-19 dataset. Specifically, through our analysis, we aim to answer the following three research questions:

\begin{itemize}
    \item \textbf{RQ1}: What are the new antisocial lexicon introduced amid the COVID-19 pandemic?
    \item \textbf{RQ2}: Who are the targeted individuals and groups of antisocial behaviors amid the pandemic?
    \item \textbf{RQ3}: What are the factors influencing the generation of antisocial content generated?
\end{itemize}

\subsection{Annotation Case Study}
Before performing the empirical analysis, we first conduct some case studies to examine the quality of our antisocial behavior annotation. Table~\ref{tbl:lex-API-comparison} presents several examples randomly sampled from our annotated COVID-19 datasets.

\begin{table}[t!]
\caption{Ten sampled tweets and their annotated label based onthe  lexicon-based method and the \textit{Perspective} API.}
\label{tbl:lex-API-comparison}
\centering
\begin{tabular}{|c|p{6.3cm}|c|c|}
\hline
\textbf{ID} & \textbf{Tweet text} & \textbf{Lexicon-based} & \textbf{Perspective API}   \\ \hline\hline
1 & Her coochie probably got the cure for corona in it. & Antisocial & Normal\\\hline
2 & RT @USER Had selfish \#China notified the world in time, the \#ChineseVirus would have died by now. But don't worry world will as always come out of this Chinese Problem as well. \#ChineseVirus \#coronavirus \#COVID-19 \#IndiaFightsCorona & Antisocial & Normal\\\hline
3 & Donald John Trump, the greatest narcissist in the history of humanity \#donaldtrump \#narcissist \#WHO & Antisocial& Normal\\\hline
4 & Hoping heat kills \#coronavirus!! & Normal & Antisocial \\\hline
5 & RT @USER: Earlier: Islam has nothing to do with terrorism  Now: \#China has nothing to do with Corona Virus.  \#ChineseBioterrorism \#COVID2019 & Normal & Antisocial \\\hline
6 & On this \#AprilFoolsDay don't go out... It's a lockdown you Indians! You can't fool \#Corona. & Normal & Antisocial\\\hline
7 & Support Social Distancing; in fact as a photographer, I have turned down jobs for my safety and for all! Let us all support \#SocialDistancing \#StayHomeSaveLives Let us join hands together and fight this monster called \#COVID-19 Yes \#WeWillWin & Antisocial & Normal \\\hline
8 & RT @USER: These online assignments will kill me way before the corona. & Normal & Antisocial \\\hline
9 & Can this corona virus get done so my man can see his pet rats again and stay the night with me so I can be held and love all over him? & Antisocial & Normal \\\hline
10 & \#COVID is that childhood loser who refuses to grow up and seeks revenge in old age. & Antisocial & Antisocial \\\hline
\end{tabular}
\end{table}

We could only compare the annotated labels with the two methods as we do not have the ground truth labels of the tweets. From Table \ref{tbl:lex-API-comparison}, we observed that there are scenarios where a lexicon-based method provides more reasonable annotation than \textit{Perspective} API and vice versa. For instance, the annotations for tweets (1)-(4) by the lexicon-based method seem more reasonable compared to those annotated by the \textit{Perspective} API. We hypothesize that the \textit{Perspective} API's inappropriate annotation on the four tweets stems from the insufficient training examples of rare occurrence words (e.g., the word ``coochie'' in tweet (1)). Conversely, the \textit{Perspective} API might treat normal tweets with potentially inappropriate keywords as antisocial content, even though it might be the keyword is correctly and appropriately used in the specific context (e.g., the word ``kill'' in tweet (4)).

There are also situations where the \textit{Perspective} API is able to provide more reasonable annotation. For example, for tweets (5)-(7), the \textit{Perspective} API seems to give more suitable labels. Specifically, for tweets (5) and (6), the lexicon-based method cannot detect the antisocial content in the tweet as there are no matching antisocial keywords found in the tweets. Nevertheless, the \textit{Perspective} API is able to overcome this limitation to provide a more appropriate label.


While the lexicon-based method and the \textit{Perspective} API collectively provided reasonable antisocial behavior annotations on our COVID-19 dataset, some false positives can still be observed in the annotated dataset. For instance, tweet (8)-(10) seems to be normal tweets. However, due to our annotation strategy, these tweets will be falsely annotated as  ``antisocial'' as one of the two methods wrongly labeled the tweets as such. In particular, tweet (8) is annotated as ``antisocial'' by the  \textit{Perspective} API. A possible reason for the annotation could be due to the negative sentiment in the tweet. In tweet (9), the lexicon-based model annotated it as antisocial based on the matching keyword ``rats'' in the \textit{extended lexicon}. Similarly, tweet (10) contains the matching keyword ``loser''; hence the lexicon-based model labeled it as ``antisocial''. These examples highlight the limitation of our automatic annotation framework. For future work, we will explore more annotation methods to improve the quality of the annotation in our COVID-19 dataset.

\subsection{New Antisocial Lexicon Amid COVID-19}

\begin{figure*}[t!]
\begin{subfigure}{.5\textwidth}
    \centering
	\includegraphics[width=0.98\textwidth]{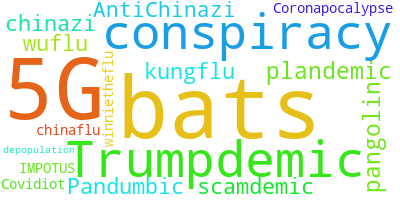}
	\caption{Unigram keywords.}
	\label{fig:AS_keywords-uni}
\end{subfigure}
\begin{subfigure}{.5\textwidth}
    \centering
	\includegraphics[width=0.98\textwidth]{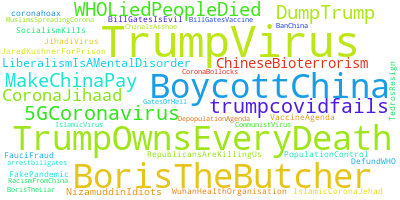}
	\caption{Bi-, tri- and four-gram keywords.}
	\label{fig:AS_keywords-more}
\end{subfigure}
\caption{High frequency antisocial keywords}
\label{fig:AS_keywords}
\end{figure*}

To answer the research question \textbf{R1}, we first perform word frequency count operation on the tweets annotated as ``antisocial'' content. The goal is to examine the popular keywords used in antisocial tweets. Fig. \ref{fig:AS_keywords-uni}) and \ref{fig:AS_keywords-more} show the high frequency unigram and multi-gram antisocial keywords found in our COVID-19 dataset respectively. We observe a number of China-related antisocial keywords such as ``wuflu'' (combination of ``Wuhan''and ``flu''), ``kungflu'', ``chinazi'', ``ChineseBioterrorism'', ``BoycottChina'', 
``HoldChinaAccountable'' etc. This supports earlier study \cite{schild2020go}, which suggests a strong presence of sinophobic behavior amid the COVID-19 pandemic. Interestingly, we also observe antisocial keywords targeting other individuals and groups. For example, we notice several keywords such as ``TrumpVirus'', ``Trumpdemic'', ``TrumpOwnsEveryDeath'', ``TrumpGenocide'', ``TrumpIsAnIdiot'', ``TrumpPandemic
''etc., which targeted at United States President Donald Trump. Similar observations are also made for antisocial keywords targeting British Prime Minister Boris Johnson. Additionally, some antisocial keywords reflect people perception about pandemic (e.g., ``Coronapocalypse'', ``scamdemic'',``plandemic'', etc.). The observations made on the high-frequency antisocial words suggest that there might be other antisocial behaviors amid the COVID-19 pandemic besides the sinophobic behavior presented in \cite{schild2020go}. We will further examine the potential antisocial content targets in the next subsection. 

Interestingly, we also observed that many of the high-frequency antisocial keywords are new terms that are not found in the open-source traditional antisocial content lexicon. This suggests that new antisocial keywords are created amid the COVID-19 pandemic. This observation also further highlights the limitations of applying the lexicon-based annotation method on fast-evolving social media datasets. Therefore,  more research will need to be done to improve the antisocial behavior annotation and detection methods.



\subsection{Antisocial Target Individuals and Groups}
From the antisocial lexicon analysis, we notice the introduction of new antisocial keywords that are targeted on specific individuals and groups. To further verify the targets of antisocial behaviors amid the COVID-19 pandemic (\textbf{R2}), we first train a word2vec model \cite{mikolov2013distributed} our annotated antisocial behavior COVID-19 dataset. Next, we query the trained wordvec model with potential target individuals and groups keywords and search for their neighboring words in the word vector. The intuition is that words that are closer to the target individuals and group keywords are either semantically close to the target or frequently used with the target. Finally, we examine these neighboring words and identify the keywords that are more frequently used in antisocial tweets.

\begin{figure}[t!]
    \centering
	\includegraphics[width=0.85\textwidth]{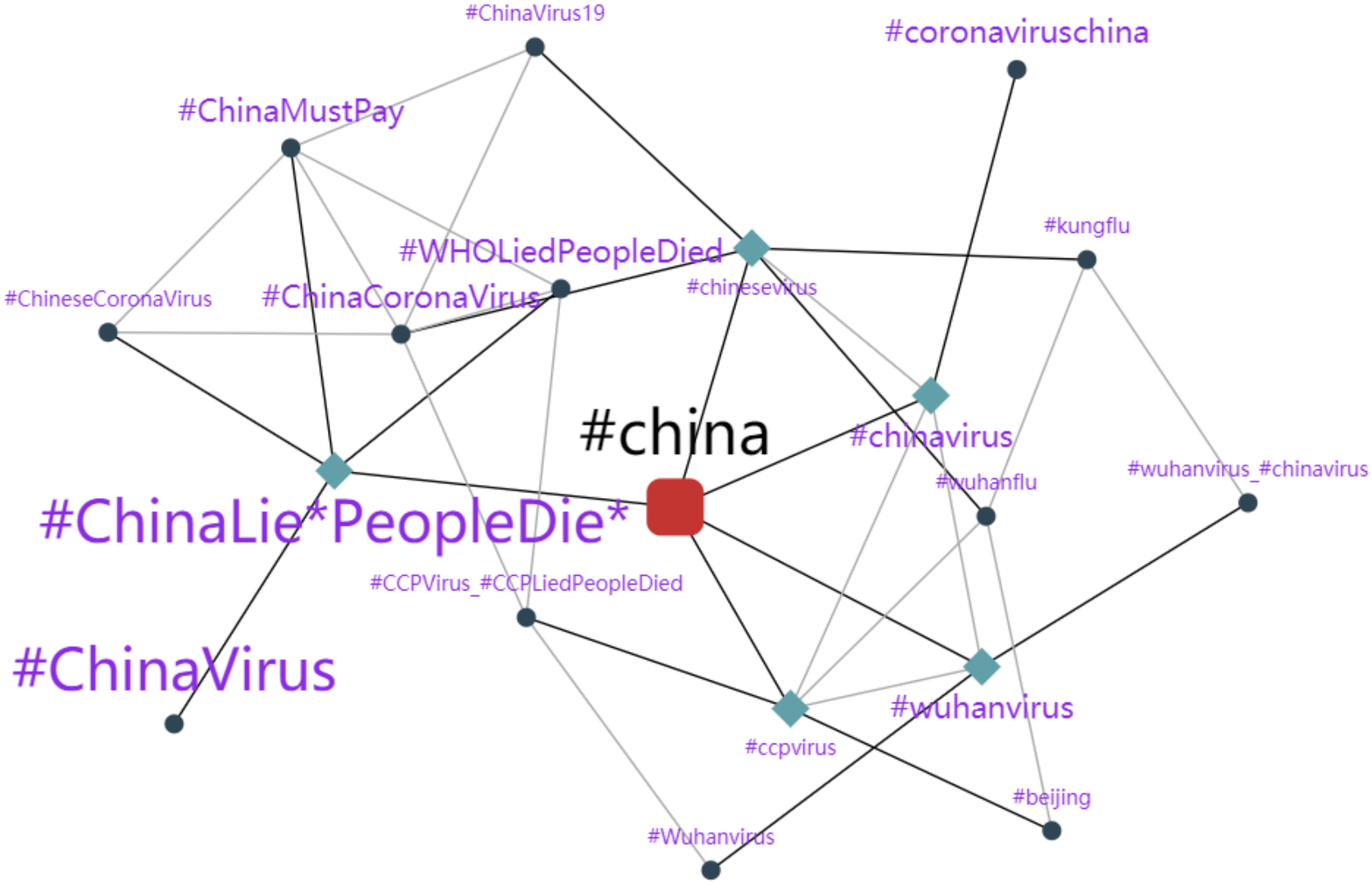}
\caption{Neighboring antisocial keywords for the target word ``China''.}
\label{fig:China_graph}
\end{figure}

Motivated by the earlier study on sinophobia amid COVID-19 pandemic \cite{schild2020go}, we first check the neighboring keywords of the target group ``China''. Fig. \ref{fig:China_graph} illustrates the graph of neighboring antisocial keywords for the target group ``China''. The target word is represented with a red square. The first-order neighbor keywords are marked with diamond symbols and second-order with circles. The 
distance of the edges corresponds to the proximity (or similarity) of the terms in vector space. Our study supports the finding in \cite{schild2020go}; we observe new antisocial keywords created that discriminate against China and the Chinese community. As the Coronavirus was assumed to have originated from the Wuhan city in China, we observed that the virus was not only being referred to as ``China virus'' or ``Chinese virus'', but new blame-attributing and conspiracy lexicon were used on China and the Chinese community. For example,  ``\#ChinaLiedPeopleDied'', ``\#ChinaMustPay'', ``\#ChineseBioterrorism'', 
``\#BanChina'', etc.

\begin{figure}[t!]
    \centering
	\includegraphics[width=1\textwidth]{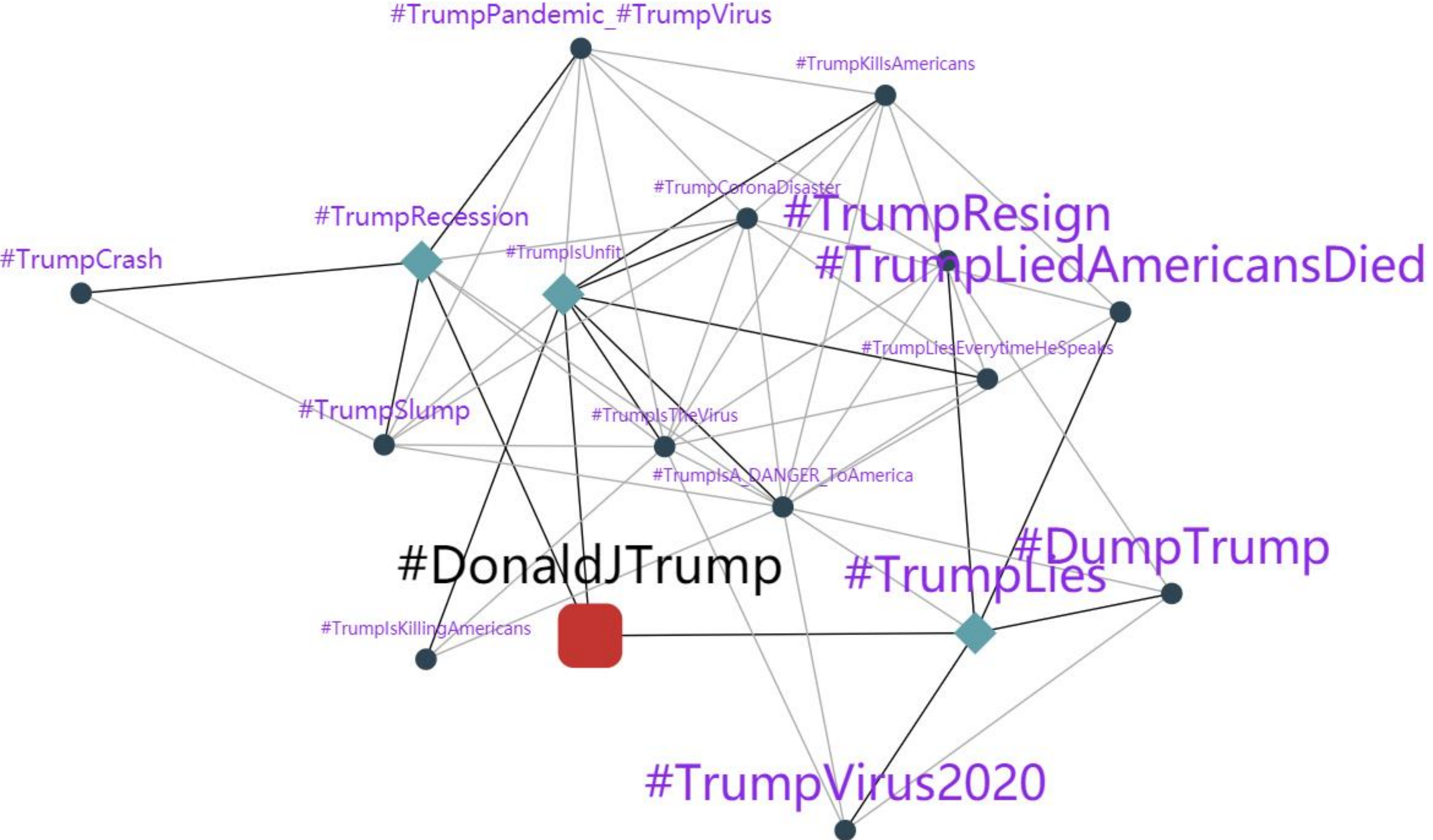}
\caption{Neighboring antisocial keywords for the target word ``DonaldJTrump''.}
\label{fig:Trump_graph}
\end{figure}

\begin{figure}[h!]
    \centering
	\includegraphics[width=0.8\textwidth]{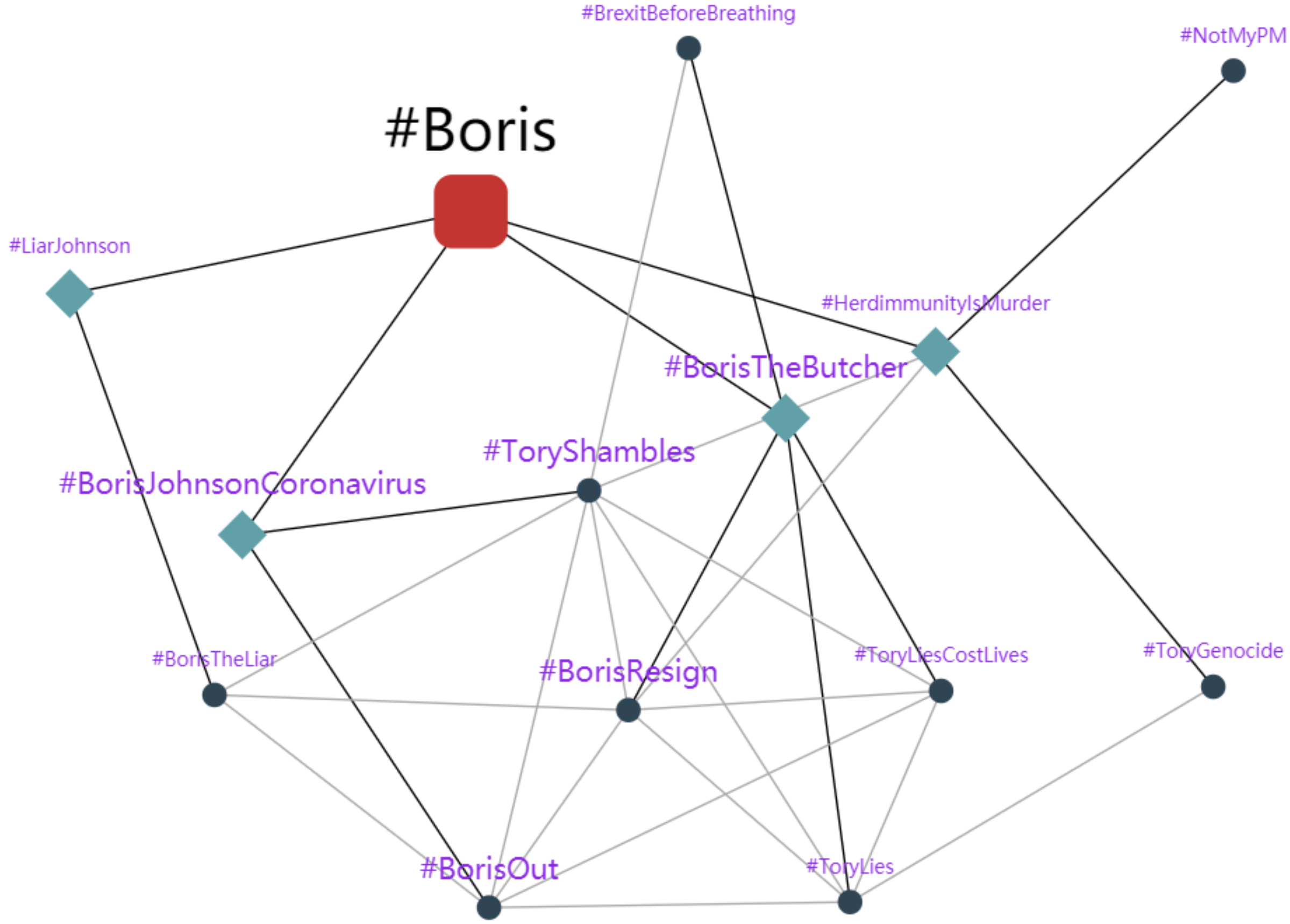}
\caption{Neighboring antisocial keywords for the target word ``Boris''.}
\label{fig:Boris_graph}
\end{figure}

While we observed intensive antisocial behaviors against China and the Chinese community, the antisocial behaviors generated amid the COVID-19 pandemic is more than just sinophobia. Prominent politicians and global NGO such as World Health Organization (WHO), are also targets of antisocial behaviors. Fig. \ref{fig:Trump_graph} and \ref{fig:Boris_graph} show the graph of neighboring antisocial keywords for the prominent politicians, United States President Donald J Trump (``DonaldJTrump''), and British Prime Minister Boris Johnson (``Boris''), respectively. We observed abusive terms such as ``TrumpVirus2020'', ``TrumpPademic'', ``DumpTrump'' etc., are frequently used on Donald Trump. Similar abusive keywords are also used on Boris Johnson (e.g., ``BorisTheButcher'', 
``ToryShambles'', etc.). Regardless of political affiliations and agendas, we believe that no individuals and groups should be subjected to antisocial behaviors in online social media.      

Many other prominent individuals are targets of baseless conspiracies and abusive tweets amid the COVID-19 pandemic. For example, prominent businessman Bill Gates (``\#BillGatesIsEvil'', ``\#GatesOfHell'', ``\#arrestbillgates'', ``\#VaccineAgenda'', etc.), immunologist Dr. Anthony Fauci (``\#FauciFraud'', etc.), and Dr. Tedros Adhanom from World Health Organization (``\#TedrosLiedPeopleDied'', ``\#WHOLiedPeopleDied'' etc.). Other races that previously had been subjected to intensive discrimination and racism \cite{schmidt2017survey,fortuna2018survey} were also targeted during the COVID-19 pandemic. For example, ``\#MuslimsSpreadingCorona'', ``\#IslamicCoronaJehad'', ``\#NizzamudinIdiots'', ``\#MuslimVirus'', ``\#CoronaJihaad'', ``\#JihadiVirus'', etc., the racist and abusive terms as such are used on the Muslim community.

To summarize, we observed antisocial behaviors targeting a wide-range of individuals and groups. Some of the targets are unique to the COVID-19 pandemic, while some groups that were previously subjected to discrimination and racism are also targeted in the COVID-19 context. Such toxic behaviors harm the society's cohesion and deepen the divides among the communities and social groups. Therefore, it is important to develop solutions to detect, curb, and monitor such undesirable online behaviors.

\subsection{Factors Influencing the Spread of Antisocial Content}
The question of what are the factors influencing the spread of antisocial content (\textbf{R3}) is a difficult problem as there could be many factors that affect the diffusion of content over social media \cite{li2017survey}. We attempt to provide a preliminary analysis of this problem by examining the temporal distribution of antisocial content generated in our COVID-19 dataset. Specifically, we compute the proportion of antisocial tweets on a given day over the observed period in our dataset. 

\begin{figure}[t!]
    \centering
	\includegraphics[width=1\textwidth]{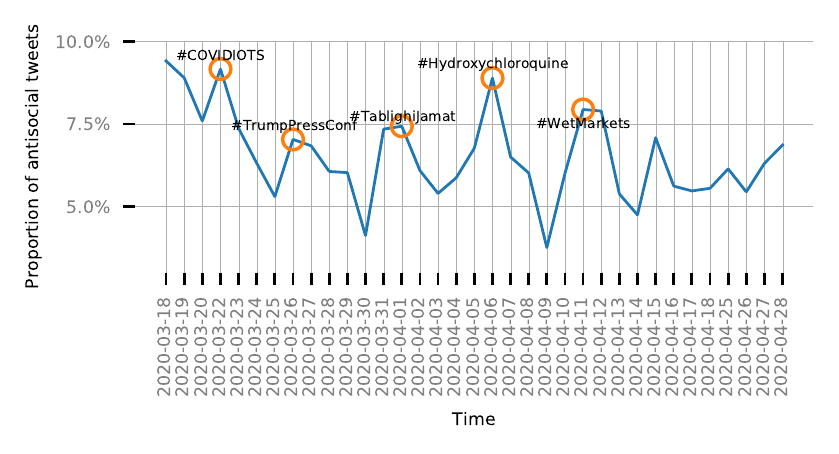}
\caption{Temporal distribution of antisocial tweets amid COVID-19 pandemic}
\label{fig:temporal}
\end{figure}

Fig. \ref{fig:temporal} shows the temporal distribution of antisocial tweets. We observe that the proportion of antisocial tweets generated per day ranges from 4\% to 9\%. There are also certain days where we notice a spike in antisocial behaviors. To understand what causes these sudden sharp increases in antisocial behavior, we dig deeper and examine the tweets on days where we observe the spike in antisocial content. Interestingly, the increase of antisocial tweets on 26th March is largely criticizing Donald Trump's press conference on the COVID-19 pandemic. The sharpest increase of antisocial behavior is observed on 6th April. Examining the tweets, we found that most of the abusive tweets are protesting against Donald Trump's claims that the medicine hydroxychloroquine cures the coronavirus\footnote{https://www.theguardian.com/global/video/2020/apr/06/trump-grilled-over-continued-promotion-of-hydroxychloroquine-to-treat-coronavirus-video}. Not all antisocial tweets are from or about the COVID-19 situation in the United States. The spike of antisocial tweets on 1st April was attributed to the criticism of the religious gathering by the Tablighi Jamaat, claiming that the event increased the spreading of coronavirus in India\footnote{https://www.bbc.com/news/world-asia-india-52131338}.

Nevertheless, it is challenging to explicitly attribute all spikes of antisocial content to a certain event. There could be many factors affecting the spread of antisocial content, and these relationships between these factors may also have an impact on the diffusion of antisocial content. For instance, we notice that the retweet function in Twitter plays a profound role in the diffusion of antisocial content. For example, when examining the spike of antisocial tweets on 11th April, we observed that many of the antisocial tweets are retweets of Bill Maher's discriminatory tweet: ``\textit{China is a dictatorship that, for decades, enforced a one child per family policy under penalty of forced sterilization. But they can't close down the farmer's market from hell? \#CoronaVirus \#WetMarkets}''. 

Our preliminary analysis of antisocial tweets exposes the complexity of antisocial content diffusion in social media.  More in-depth research will have to be conducted to curb the spread of these toxic behaviors to analyze the multiple factors that affect the spread of online antisocial behaviors.

\section{Conclusion and Future Works}
\label{sec:conclusion}
Online antisocial behavior coarsens public discourse and weakens the social fabric. The presence of such toxic behaviors takes a heavy toll on the already daunting COVID-19 global pandemic. Despite the gravity of the issue, very few studies have studied online antisocial behaviors amid the COVID-19 pandemic. In this paper, we filled the research gap by collecting and annotating a large dataset of over 40 million COVID-19 related tweets. Specially, we designed an annotation framework that combines a lexicon-based method and the \textit{Perspective} API to annotate the antisocial behavior tweets automatically. We performed empirical analysis on our annotated dataset. Our study found that new abusive lexicons are introduced amid the COVID-19 pandemic. We also empirically identified the vulnerable targets of antisocial behaviors and some of the factors that influenced the spreading of online antisocial content during the pandemic.

Our study provides a preliminary analysis of the online antisocial behavior amid COVID-19. More future works will still need to be done to tackle the pressing problem. For instance, better methods need to be developed to annotate the antisocial content in the fast-evolving social media. More in-depth research will need to be conducted to learn the factors that influence the spread of online antisocial behaviors. We have released our annotated antisocial behavior COVID-19 dataset, hoping that it encourages more research in this critical field. 

\bibliographystyle{splncs04}
\bibliography{ref}

\end{document}